# The Cost of Pollution in the Upper Atoyac River Basin: A Systematic Review

María Eugenia Ibarrarán, Romeo A. Saldaña-Vázquez and Tamara Pérez-García

Instituto de Investigaciones en Medio Ambiente Xavier Gorostiaga S.J. Universidad Iberoamericana Puebla, Blvd. del Niño Poblano No. 2901, Col. Reserva Territorial Atlixcáyotl, San Andrés Cholula, Puebla C. P. 72820, México

**Abstract**

The Atoyac River is among the two most polluted in Mexico. Water quality in the Upper Atoyac River Basin (UARB) has been devastated by industrial and municipal wastewater, as well as from effluents from local dwellers, that go through little to no treatment, affecting health, production, ecosystems and property value. We did a systematic review and mapping of the costs that pollution imposes on different sectors and localities in the UARB, and initially found 358 studies, of which 17 were of our particular interest. We focus on estimating the cost of pollution through different valuation methods such as averted costs, hedonic pricing, and contingent valuation, and for that we only use 10 studies. Costs range from less than a million to over $16 million dollars a year, depending on the sector, with agriculture, industry and tourism yielding the highest costs. This exercise is the first of its kind in the UARB that maps costs for sectors and localities affected, and sheds light on the need of additional research to estimate the total cost of pollution throughout the basin. This information may help design further research needs in the region.

**Introduction**

The Upper Atoyac River Basin (UARB), located in the states of Tlaxcala and Puebla in Mexico, is among the two most polluted basins in the country (CONAGUA, 2016b; DOF, 2011; Sandoval et al, 2009; Montero et al 2006). This has been widely documented and the main causes that have been identified are industrial and municipal wastewater that often goes untreated directly to the Atoyac River and its tributaries, rivers Zahuapan and Alseseca (DOF, 2011; IMTA, 2005). This has caused awareness among the more than two million people living in the basin, as it is a threat to their safety and livelihoods (CNDH, 2017).

Pollution, and water pollution, threatens environmental quality and therefore ecosystems and human health. This affects productivity, both of ecosystems and the economy, and reduces local sources of wellbeing. Deterioration of habitats due to pollution impinges on sustainable economic growth and thus social development. At the community level, contamination leads to severe health-related issues as cancer, leukemia, and kidney problems, among others. Finally, pollution affects property value.

Additionally, to a deteriorating water quality, water consumption in the basin has increased due to domestic, agriculture, and industrial use, leading to a higher pressure for water and to overexploitation of aquifers (Pérez et al, 2018, Bravo-Cadena et al. 2021). Urban population growth has led to increased demand for water and a higher outlet of wastewater (Rodríguez, Morales & Zavala, 2012). Moreover, the sprawl and dissemination of rural population has made it increasingly difficult to provide water and sanitation, and the scarce water is overused and polluted. This had led to irrigation with highly polluted water because the lack of options (Pérez Castresana et al, 2019).

In the literature there are several types of studies of river basins. Most characterize the type of pollutants, their origin, and their effects. A few look at the costs such pollution imposes. These are precisely the ones we consider for our analysis. The aim of this paper is, therefore, to make a systematic review and mapping of literature related with the value of water deterioration in the UARB. We look at the costs in terms of disease as well as losses in economic activities and to property itself. We also review some papers on the costs of cleanup. This is only a subset of the total values that can be calculated for the region; however, it gives a lower bound for the size of the economic losses and costs produced by such pollution. In so doing, we highlight the monetary values attached to pollution, so society and policymakers can make sense of the magnitude of the impacts of such deterioration.

A few studies have been found that estimate the monetary value of pollution in different parts of the basin, particularly close to rivers. Producing more studies is expensive and probably reiterative, so we review these studies to compare values across them. From this analysis we obtain a range of values for the costs imposed by deterioration of the basin, and hope these figures are useful for policy making towards a more thorough protection of the basin.

This paper is divided into four sections. Section 1 presents the study area. Section 2 sets the framework for the analysis, and describes the information found for the basin, the methodologies used, and their relation to the environmental valuation literature. Section 3 presents the setup for the systematic review and the results. Section 4 addresses some conclusions and suggests further research.

1. **The study area**

The Upper Atoyac River Basin (UARB) is in Central Mexico, as is shown in the first part of Figure 1. It covers 69 municipalities in total, 47 in Tlaxcala and 22 in Puebla, represented in the second panel. The most relevant tributaries in the basin, shown in the final panel, are the Zahuapan, Atoyac, and Alseseca rivers, that end up in a dam south of Puebla City, in Valsequillo.

Figure 1. The Upper Atoyac River Basin

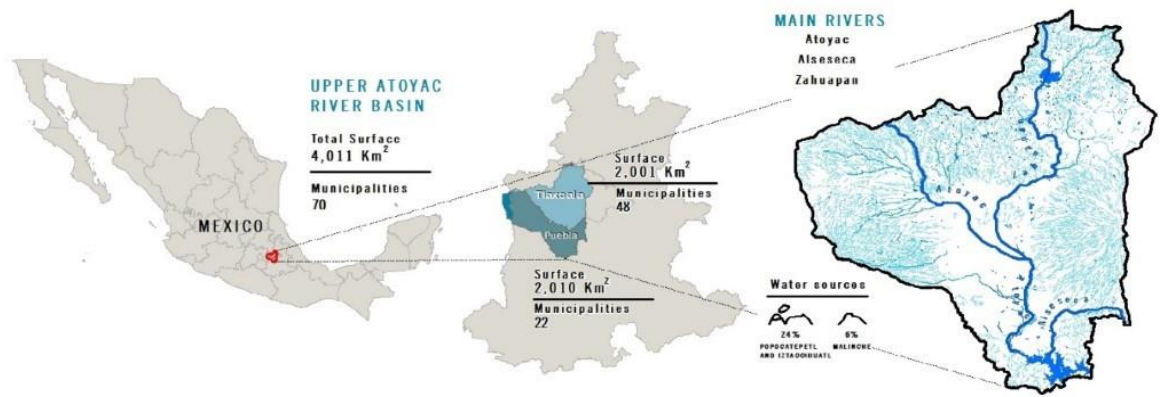

*Source*: Own.

With about 3 million inhabitants, the Puebla-Tlaxcala Metropolitan Area is the fourth largest nationwide in terms of population, and with a high pressure on water resources (CONAGUA 2016a; CONAGUA, 2016b; Rodríguez & Morales 2014; DOF, 2010).

Additionally, there is a high persistence of water toxicity, affecting its communities (Pérez et al, 2018; Martínez et al, 2017). Furthermore, water from this basin reaches the Manuel Ávila Camacho dam, which is then used to irrigate the agricultural district of Valsequillo, before it follows its course into other states. This represents a risk for food quality in terms of organic pollution and heavy metals (Rodríguez-Espinosa et al, 2018). Problems like water scarcity and high levels of contamination have been identified in this district. Concentrations vary depending on the season, so during dry spells pollution is even worse than during the rainy season, affecting predominantly children (Castresana et al, 2019; Martínez et al, 2017; CNDH, 2017).

Regardless a vast number of studies, both locally and nationally, to diagnose the situation of water pollution, as well as many government programs to solve it and money spent in the process, pollution has not been abated, and remains a very pressing problem. Technical solutions exist, but the major concern seems to be water governance. Even though regulatory reforms have taken place in Mexico since 2000, reports state there have not been significant changes regarding water governance either regionally (Ocampo-Feletes, Parra-Inzunza & Ruiz-Barbosa, 2018; Casiano, Bressers & Gleason, 2017; Casiano & Bressers, 2015; Rodríguez & Beristain, 2014; Sierra, Pale & Serrano, 2005), or at the national level.

## 2. Framework of analysis

We now turn to describe the framework of analysis that we use for our study. Then we address the existing literature and fit it to such a framework. Using that literature, we develop the valuation review in the following section.

We use the framework of environmental economics to describe the value of the environment, and valuation methods to assess the harm impinged on it. Values are difficult to quantify because usually there are no markets and therefore no prices attached. Therefore, the goal of this research is then to quantify the value being destroyed by pollution.

The environment has use and non-use values. Use values can be addressed by the benefits we derive from its use through direct or indirect consumption or production processes. Non-use values are derived from the sheer fact that it exists, because we think they may be valuable to other generations, or just because knowing it is there provides joy. The value of existence, for example, is a non-use value. Together use and non-use values give the total value of an environmental good or service (Bastien-Olvera & Moore, 2020;

Lara-Pulido et al., 2018; Knowler, 2005; Field & Field, 2003; Pearce et al, 1999; Hearne, 1996).

Use values can be determined directly by asking people how much they are willing to pay for a certain environmental good or service. This direct valuation method is known as contingent valuation (Mitchell & Carson, 1989). However, often the value of environmental goods and services must be determined indirectly, by looking at the expenditure on private goods that may be substitutes or complements of environmental goods. Such expenditure may work as proxies for the value of environmental goods. Valuation methods have been developed to determine these values. When private and environmental goods are substitutes, the averted or induced costs method may be used. When they are complements, the travel cost method helps infer the value of the environmental good. Finally, the value of the environmental component of a private good may be determined using the method of hedonic pricing (Popa & Guillermin, 2017; Dixon et al., 2013; Field & Field 2003; Azqueta, 2007). On the other hand, non-use values may also be obtained using the contingent valuation method described above. Figure 2 presents this framework for the study region. Our literature review for the UARB has identified studies estimating mostly use values. We have found studies on the effects of water pollution on health, different economic sectors, the value of property, and on ecosystems. The figure shows the valuation methods used in each case, i.e., induced costs, hedonic pricing, and contingent valuation.

Figure 2. Valuation framework for the Upper Atoyac River Basin

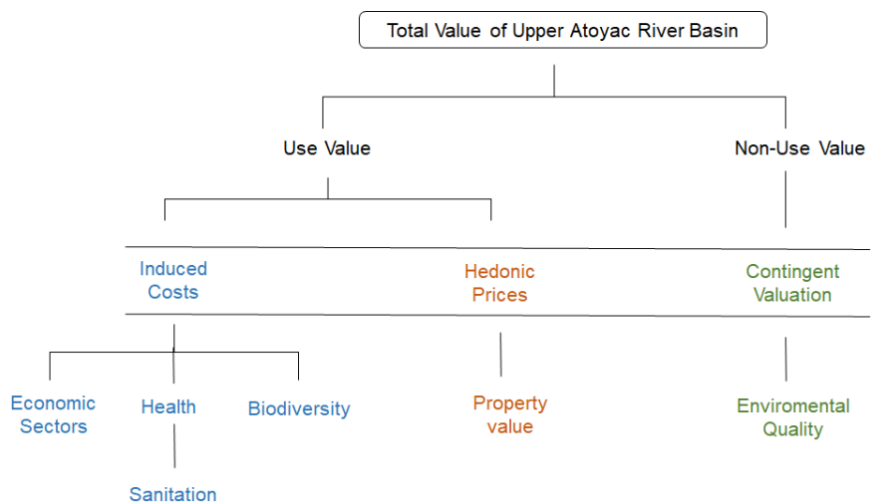

*Source:* Own.

The quality of river basins as the one we want to value here are quite complex in that they depend on the integrity of other ecosystems such as forests, the care of the environment that large industrial and urban areas offer, and on the protection authorities give through regulation and its enforcement. Governance of the basin is crucial, since the interaction of parties leads to a certain level of water security, defined by the United Nations as "the capacity of a population to safeguard sustainable access to adequate quantities of acceptable quality water for sustaining livelihoods, human well-being, and socio-economic development, for ensuring protection against water-borne pollution and water-related disasters, and for preserving ecosystems in a climate of peace and political stability" (UN Water, 2013). Water security, therefore, summarizes the quality of interaction of parties within the basin.

In general, the quality of a river basin is a *public good*, where there is non-rivalry in its consumption and non-excludability. Its degradation, on the other hand, is a public bad, also non-rival in consumption and very costly to exclude oneself from its dangers. Precisely these two characteristics of *public bads* are what make pollution of the basin so detrimental to the communities living there: pollution may affect all, regardless that others are affected, and it is very costly or literally impossible to protect from it (Atkinson & Stiglitz, 2015; Stiglitz & Rosengard, 2015).

Even though these adverse effects have been documented and even used for lawsuits against polluters and government officers not taking charge, such lawsuits have been based on the effect pollution produces, basically on health (CNHD, 2017). The literature on pollution, its sources, concentrations, and effects, is therefore vast (Rodríguez-Tapia & Morales-Novelo, 2017).

There are few efforts to quantify, in monetary terms, the effects of such degradation on productivity, namely on agriculture, livestock and fisheries, and on tourism, or the induced costs for communities that must find other sources of water or whose property value collapses (Rodríguez-Tapia et a., 2012).

On the other hand, these studies cover segments of the basin, usually transects of the Atoyac River. There are parts of the basin that have been studied at depth such as Atoyac and the Zahuapan sub-basins (de Oca & Fuentes, 2019 and 2017; Rodríguez Tapia, et al., 2012; Jiménez & Hernández, 2011). Other regions have been largely ignored, as is the case of the Alseseca sub-basin. A thorough and systematic integration of the existing information is called for to address the overall costs imposed on the basin by its deterioration, and to identify where studies are still needed.

3.  **Literature search protocol and results**

The literature review was done using Google Scholar, because this academic research engine includes studies published in journals, thesis, and projects reports ("grey literature"), making our literature review broader (Haddaway et al., 2015). The keywords used were "valoración económica AND río Atoyac" (search date: May 7, 2020), "daños por contaminación AND río Atoyac" (search date: May 11, 2020), "economic value AND Atoyac River" (search date: May 11, 2020), and "pollution damage AND Atoyac river" (search date: May 12, 2020). These words were searched in the titles and the body of the documents. There was no constraint imposed based on year of publication. Additionally, manual searches were done in the documental repository of the Atoyac River held in the Xabier Gorostiaga Environmental Research Institute at Universidad Iberoamericana Puebla (IIMA, due to Spanish acronym), and dissertations published by our research group were also reviewed. This repository was built by including resources found along the past five years in journals, reports, thesis, and government studies (search date: May 19, 2020). Many documents were provided by Dale la Cara al Atoyac, a local NGO.

Figure 3. Most frequent topics found in the reviewed studies

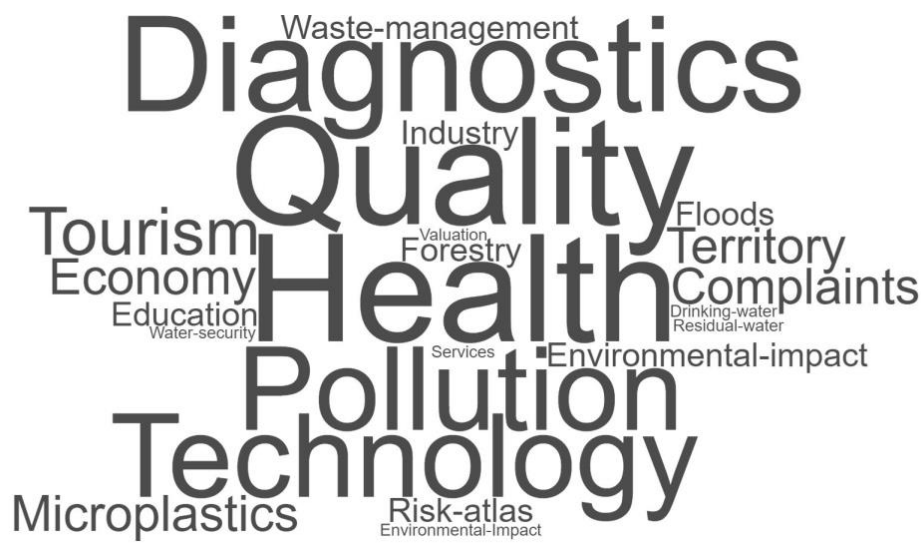

*Source*: Own.

From the literature review for the UARB, we found 358 studies. We classified the topics of studies revised based on the word cloud below. Most of the studies refer to water quality, health, pollution, and diagnosis of the region, as well as technology available to clean the river, as shown in Figure 3. Valuation, in very small print, is evidently quite low in this word cloud diagram.

*Studies selection criteria*

Once the documents were obtained, title and abstracts of the selected papers were revised to assure that they followed the requirements so that the proper information could be extracted, namely, the economic valuation associated to the pollution in the UARB. The inclusion criteria to select these documents were that (1) the studies were carried out in the UARB; (2) they contain an economic valuation of the effects of pollution of the UARB; and (3) the studies had an original source of data. This allowed us to avoid duplicating the cost of pollution in the UARB.

*Search process and studies used*

The economic valuation studies on the effects of pollution of the Atoyac River are scarce. Figure 4 shows the flow diagram from the search process according to the PRISMA convention for the systematic revision of literature and meta-analysis (Urrútia & Bonfill, 2010).

From the 17 studies identified, a very small set of 6 proved to be useful for the cost analysis. Their data was collected in 2005, 2007, and 2009 mainly, as shown in Table 1. We selected studies 1 through 6 because they are the primary source of cost of pollution calculations. Studies 5 through 10 are used to identify location. Studies 5 and 6 have both costs and locations.

Figure 4. Search process and results

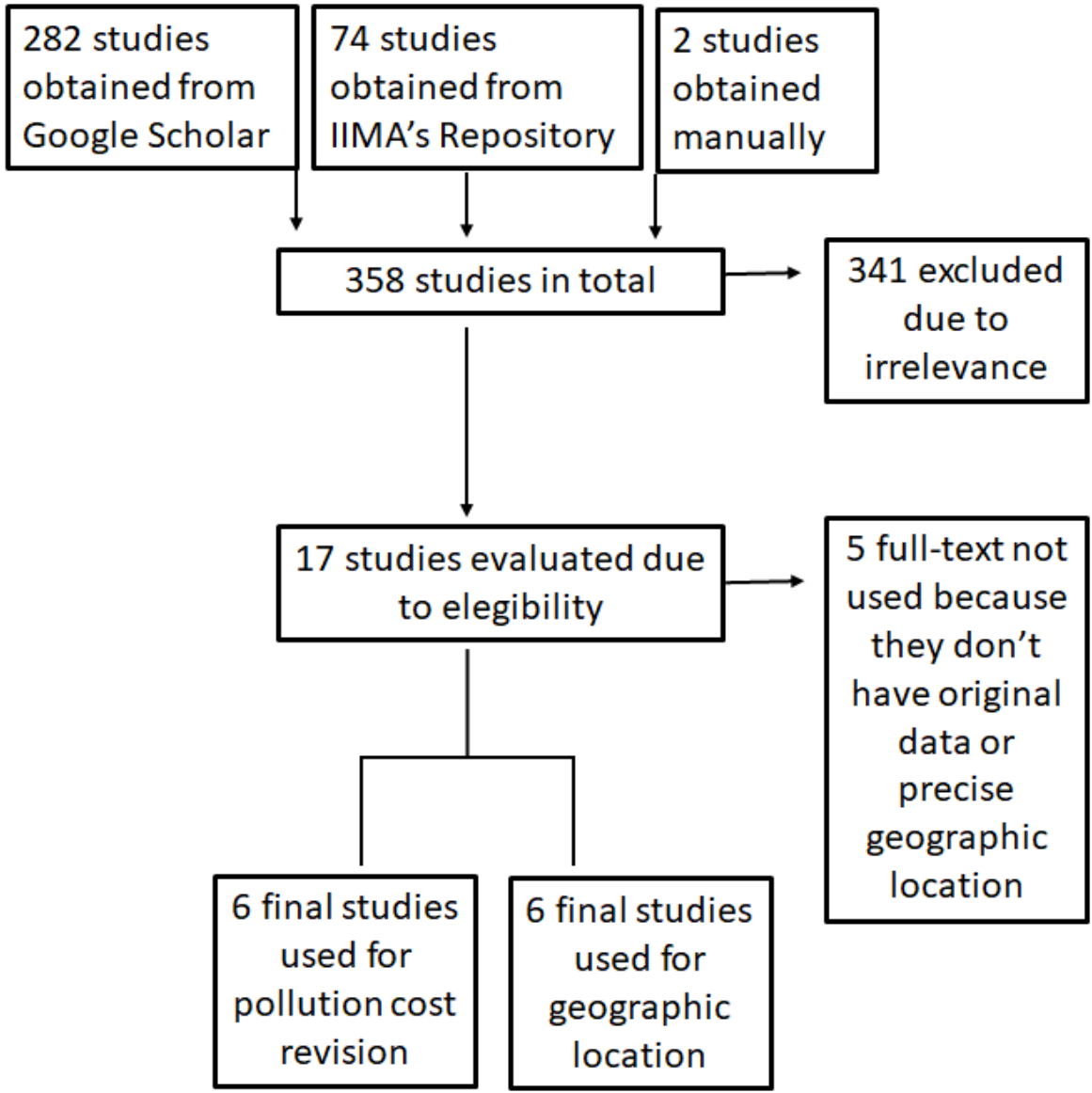

*Source*: Own.

Table 1 lists in detail the studies used for the rest of the paper. Some report use values and others non-use value, following the methods described in Figure 2.

Table 1. Eligible studies

| ID | Author, year of publication | Publication type | Year of collected data | Estimated value | | Method |
| --- | --- | --- | --- | --- | --- | --- |
| | | | | Use | Non-Use | |
| 1 | Saldívar Valdez (2007) | Report | 2007 | Econ Sect Health Biodiversity | | Induced cost |
| 2 | Saldívar Valdez & Olivera Villarroel (2011) | Book Chapter | 2007 | Health Biodiversity | | Induced cost |
| 3 | Rodríguez-Tapia et al. (2012) | Journal article | 2005 | Health Econ Sect | | Induced cost |
| 4 | Soto Montes de Oca (2013) | MA Thesis | 2009 | | Environ Quality | Contingent valuation |
| 5 | Aquino Moreno et al. (2015) | Report | 2013 | Health | | Induced cost |
| 6 | Soto Montes de Oca (2009) | Business Report | 2009 | | Environ Quality | Contingent valuation |
| 7 | Puebla State Government (2010) | Government Report | 2009 | Econ Sect | | Induced cost |
| 7 | | | | | Environ Quality | Contingent valuation |
| 8 | Gómez et al. (2002) | Report | 2002 | Econ Sect Sanitation | | Induced cost |
| 9 | Sánchez Castañeda et al (2017) | Municipal Government Report | 2005 | Property Value | | Hedonic Prices |
| 9 | | | 2009 | Health | | Induced cost |
| 9 | | | 2013 | Econ Sect | | Induced cost |
| 9 | | | 2016 | | Environ Quality | Contingent valuation |
| 10 | Headley Olivo & Salas Manzur (2019) | BA Thesis | 2016 | Econ Sect | | Induced cost |

*Source*: Own.

*Location*

We first map the location of studies that contain geographic coordinates. Studies are marked as colored circles, depending on the sector, and located throughout the region. Several studies were simultaneously done in several sites, therefore there are more locations on the map than the number of studies revised. We map studies 5, 8, and 9 that evaluate the cost of pollution on health and sanitation; 7, 8, 9, and 10 that show the cost on economic activities; 9 on property values; and 3, 6, and 7 of changes in environmental quality. As can be seen in Figure 5, most studies do not cover areas with a high population density. Those on areas with a low population density are of the primary sector and environmental quality. Studies also follow the distribution of industry, particularly textile, chemical and automotive that are the most polluting.

Figure 5. Areas covered in studies, population density, and industrial location

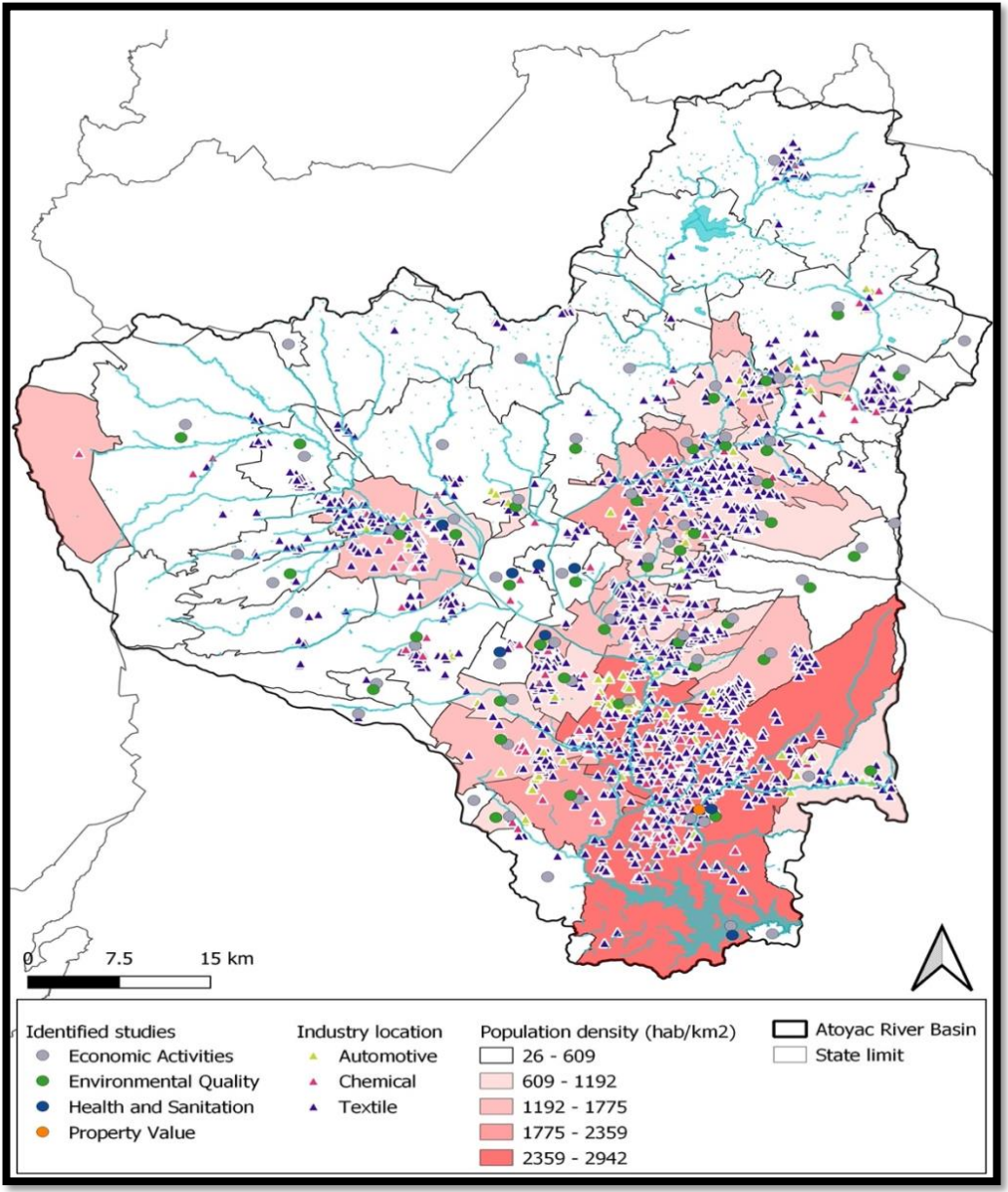

*Source*: Own, based on Sistema Estatal de Información (2015), Agenda Estadística (2016) and INEGI (2019).

*Costs*

We report two sets of costs separately. First, we look at the induced costs, i.e., costs imposed on the sectors because of the extra costs they must withstand to operate, due to a reduction of their production, or to lower prices of the goods they sell. Health has the highest reported value of $10.1 million USD/year, while other health studies report a value ranging from 10 thousand to $10 million dollars a year. Other high costs accrue to industry, that reports a point value of roughly $9.3 million, derived from the cleanup costs of their wastewater. Agriculture and tourism face costs ranging from $2 to $16.1 million dollars per year. Biodiversity faces a cost of approximately $1.9 million USD per year. Finally, livestock, fisheries and sanitation (cleanup of the river) face the lowest costs. These numbers, it is important to point out, are reported on a yearly basis for a concrete location and sector, and do not represent the full cost for the sector in the basin and or for a larger period.

Figure 6. Range of induced costs reported by sectors *(in constant 2020 US dollars)*

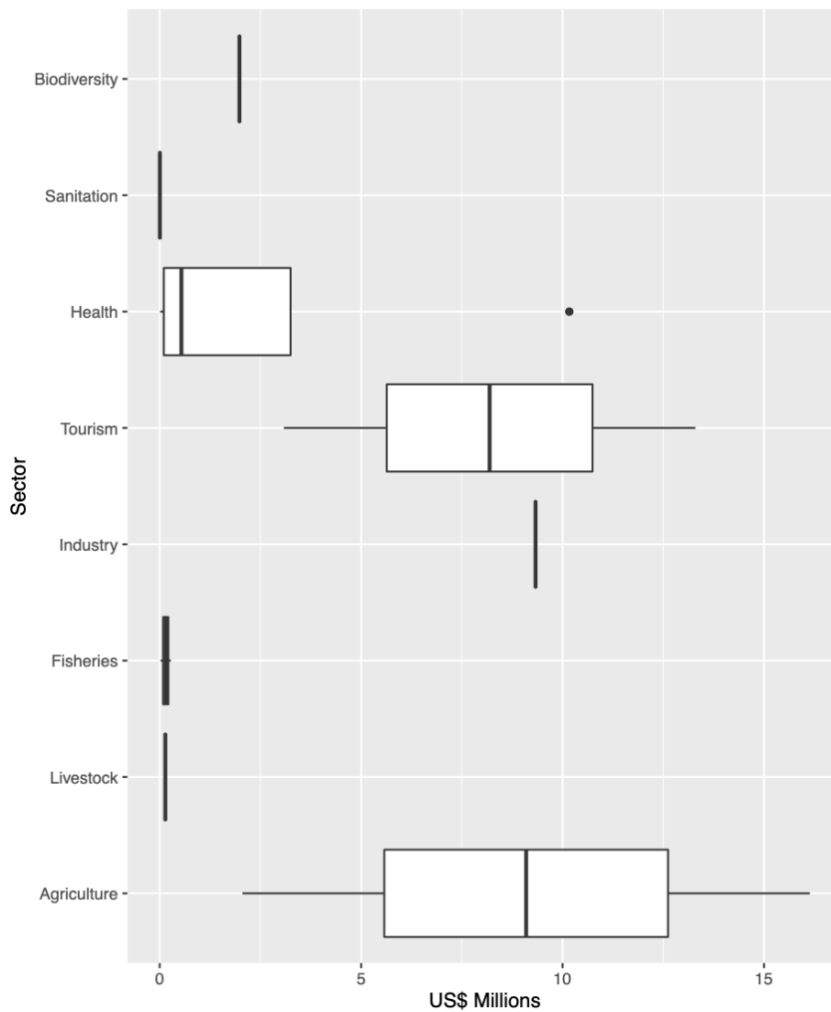

*Source*: Own, based on deflated costs presented in Appendix 2.

Costs vary significantly across economic sectors, and often within the sector. For example, within the primary sector, livestock and fisheries show a lower cost from pollution compared with agriculture. These results could be a response of the different magnitude of each activity in the basin and the relative magnitude of water use. Crop production is the primary activity that uses most of the surface water available in the UARB basin (Bravo-Cadena et al. 2021, CONAGUA 2018, CONABIO 2011). Therefore, we could infer that the cost to this activity may be greater than for other sectors.

On the other hand, surprisingly, the costs of pollution to human health, except for the outlier, is lower than costs to other sectors such as industry, tourism, and agriculture. These reduced costs could be due to three possible factors: 1) the low number of studies done where water pollution is highest (north of the basin); 2) studies include a small number of communities with a low population; and 3) the effects of water pollution are not immediate and are cumulative.

We finally turn to discuss the costs of pollution on environmental quality. These costs are usually estimated through contingent valuation studies and show the individuals´ willingness to pay to increase environmental quality. They are not aggregated for a community or for a comparable time span, so they are not directly comparable to those in Figure 6. These numbers obtained from contingent valuation studies, namely studies 4, 6, 7 and 9 listed in Table 1, are reported in Table 2. All studies give similar values, on average willingness to pay is $13.50 USD/person bimonthly.

## 4. Final comments and further research

Many studies have been developed for the UARB. This has led to the general thought that it has been over diagnosed and that now it is time to act. In some fields this may be true, but not necessarily in the economic valuation front. Even though several studies estimate the costs pollution imposes on the different sectors, this has not been done systematically or thoroughly.

This analysis has shed light on what is available, i.e., studies for selected sectors, but for specific locations, and for one-year periods. In the case of agriculture and tourism, for example, the costs for these two sectors are on a per case basis and are not aggregated across the region. Expanding the studies to cover the entire basin and multi-decade period should be fairly simple to do in upcoming studies.

The costs to the industrial sector are only found in one study and they refer to the cost of installing equipment to clean up their own wastewater. Plenty of more studies can be made for the main industrial sectors of the region, reporting the cleanup costs for the textile, automotive, and the most prominent polluters.

Economic valuation of health costs is particularly important since these must focus on chronic illnesses rather than on diarrhea. This will certainly increase the value of such costs to those of the higher bound outlier than to the lower costs where most studies are reported. Eventually, when building a thorough cost-benefit analysis of reducing pollution in the basin, the benefits of such actions could be taken from the present value of the foregone health costs, if they cover all chronic illnesses within the basin.

The costs to biodiversity and environmental quality have been estimated through two different routes, namely induced costs and contingent valuation respectively. The way these costs are reported are not comparable since the former is an aggregate cost and the latter is individual and for a shorter period. Further studies to determine the cost of pollution to the environment, be it biodiversity or ecosystems, must be designed carefully to produce adequate results that can then be added to the costs of other sectors and therefore obtain the present value of pollution to the basin. This may be achieved by using the current framework of the value of ecosystems services or nature contributions to people (Díaz et al. 2018).

Finally, and this holds for most of the literature reviewed, studies only cover one year. Efforts need to be made to estimates the costs of pollution over a longer time span, making sure to capture the cumulative effects of pollution.

The type of results needed, rather than only used to characterize the damage of pollution in a precise location and sector in the short-run, should focus on providing the building blocks to carry out a full benefit-cost analysis that could lead discussion and foster policies to protect the basin. Economic valuation of the environment may highlight the benefits of reducing the externalities of environmental degradation.

**Acknowledgements**

We thank Universidad Iberoamericana Puebla that provided the resources to finance this research.

**Appendix**

List of revised studies in the systematic literature review

**References**

Agenda Estadística (2016). Densidad de Población por municipio 2010 y 2015. Available online: http://evaluacion.tlaxcala.gob.mx/images/stories/documentos/planea/estadistica/agenda/agenda16/poblacion/3_2.htm (Accessed on 4 Dec 2020)

Atkinson, A. B., & Stiglitz, J. E. (2015). *Lectures on Public Economics: Updated Edition*. Princeton University Press.

Azqueta, D. (2007). *Introducción a la Economía Ambiental*, 2da edición. McGraw-Hill.

Bastien-Olvera, B. A., & Moore, F. C. (2020). Use and non-use value of nature and the social cost of carbon. *Nature Sustainability*, 1-8., B. A., & Moore, F. C. (2020). Use and non-use value of nature and the social cost of carbon. *Nature Sustainability*, 1-8.

Bravo-Cadena J, Pavón NP, Balvanera P, Sánchez-Rojas G, Razo-Zarate R. Water Availability–Demand Balance under Climate Change Scenarios in an Overpopulated Region of Mexico. International Journal of Environmental Research and Public Health. 2021; 18(4):1846. https://doi.org/10.3390/ijerph18041846

Casiano, C., & Bressers, H. (2015). Changes without changes. *The Puebla's Alto Atoyac sub-basin case in Mexico. Water Gov*, *1*(2), 12-16.

Casiano, C., Bressers, H., & Gleason, A. (2017). Evaluación de la gobernanza de la política de las plantas de tratamiento residual (2013-2016), en la zona centro de México: los casos de Tlaxcala-Zahuapan, Puebla-Atoyac y Presa Guadalupe en el Estado de México. In *1er Foro Internacional" Politicas públicas para el desarrollo sustentable: horizontes en el siglo XXI*.

Castresana, G. P., Roldán, E. C., García Suastegui, W. A., Morán Perales, J. L., Cruz Montalvo, A., & Silva, A. H. (2019). Evaluation of Health Risks due to Heavy Metals in a Rural Population Exposed to Atoyac River Pollution in Puebla, Mexico. *Water*, *11*(2), 277.

CNDH (2017). Recomendación No. 10/2017; Comisión Nacional de los Derechos Humanos, México City, México, 2017.

Comisión Nacional para el Conocimiento y Uso de la Biodiversidad (CONABIO). 2011. La Biodiversidad en Puebla: Estudio de Estado. México. Comisión Nacional para el Conocimiento y Uso de la Biodiversidad, Gobierno del Estado de Puebla, Benemérita Universidad Autónoma de Puebla. 440 páginas.

CONAGUA (2016a). Atlas del agua en México 2016, ed.; Comisión Nacional de Agua: México D.F., México, 2016; pp. 20–100. Available online: http://201.116.60.25/publicaciones/AAM_2016.pdf (accessed on 11 Jan 2019).

CONAGUA (2016b). Red Nacional de Monitoreo 2012–2015 (RNM), Comisión Nacional de Agua: México D.F., México. Available online: https://www.gob.mx/conagua/documentos/monitoreo-de-la-calidad-del-agua-en-mexico (accessed on 15 Feb 2019).

CONAGUA (2018). Actualización de la disposición media anual de agua en el acuífero Alto Atoyac (2901), estado de Tlaxcala. Available online: https://sigagis.conagua.gob.mx/gas1/Edos_Acuiferos_18/tlaxcala/DR_2901.pdf (accesed on 14 Dec 2020)

de Oca, G. S. M., & Ramíerz- Fuentes, A. (2017). Los beneficios económicos de la restauración de la cuenca de Atoyac en Puebla, México. *Espacialidades*, *7*(1), 65-98.

de Oca, G. S. M., & Ramirez-Fuentes, A. (2019). Valor del rescate de ríos cuando se vive cerca y lejos. La Cuenca de Atoyac en Puebla, México. *Tecnología y Ciencias del Agua*, *10*(1), 1-31.

Díaz, S., Pascual, U., Stenseke, M., Martín-López, B., Watson, R. T., Molnár, Z., ... & Shirayama, Y. (2018). Assessing nature's contributions to people. Science, 359(6373), 270-272.

Dixon, J., Scura, L., Carpenter, R., & Sherman, P. (2013). *Economic analysis of environmental impacts*. Routledge.

DOF (2010). ACUERDO Por el que se dan a conocer los Estudios Técnicos de Aguas Nacionales Superficiales de la Región Hidrológica Número 18 Balsas, México. 2010. Available online: http://dof.gob.mx/nota_detalle_popup.php?codigo=5175730 (accessed on 10 Jan 2019).

DOF (2011). Declaratoria de Clasificación de los ríos Atoyac, Xochiac o Hueyapan, y sus afluentes; Diario Oficial de la Federación, México, 2011.

Field, B.C. y M. K. Field (2003). *Economía Ambiental.* 3era Edición. McGraw Hill.

Haddaway, N. R., Collins, A. M., Coughlin, D., & Kirk, S. (2015). The role of Google Scholar in evidence reviews and its applicability to grey literature searching. *PloS one*, *10*(9), e0138237.

Hearne, R. R. (1996). Economic valuation of use and non-use values of environmental goods and services in developing countries. *Project Appraisal*, *11*(4), 255-260.

IMTA (2005). Estudio de Clasificación del Río Atoyac, Puebla-Tlaxcala; Instituto Mexicano de Tecnología del Agua - CONAGUA: México City, México.

INEGI (2019) Directorio Estadístico Nacional de Unidades Económicas. Available online: https://www.inegi.org.mx/app/descarga/ (Accessed on 10 Dec 2020).

Jiménez, G. R & Hernández R. M. L. (2011). *Zahuapan: Río-Región-Contaminación*. El Colegio de Tlaxcala A. C., San Pablo Apetatitlán, Tlaxcala: 469p.

Knowler, D. (2005). 'Short cut'techniques to incorporate environmental considerations into project appraisal: an exploration using case studies. *Journal of Environmental Planning and Management*, *48*(5), 747-770.

Lara-Pulido, J. A., Guevara-Sanginés, A., & Martelo, C. A. (2018). A meta-analysis of economic valuation of ecosystem services in Mexico. *Ecosystem Services*, *31*, 126-141.

Martínez, E.; Rodríguez, P.F.; Shruti, V.C.; Sujitha, S.; Morales, S.; Muñoz, N. (2017). Monitoring the seasonal dynamics of physicochemical parameters from Atoyac River basin (Puebla), Central Mexico: Multivariate approach. *Environ. Earth Sci.*, 76, 1–15.

Méndez, P. (December 15[th], 2020) Avalan ley de egresos del estado para 2021 con recortes para 13 dependencias. La jornada de oriente. https://www.lajornadadeoriente.com.mx/puebla/ley-de-egresos-para-2021/

Mitchell, R. C., & Carson, R. T. (1989). *Using surveys to value public goods: the contingent valuation method*. Resources for the Future.

Montero, R., Serrano, L., Araujo, A., Dávila, V., Ponce, J., Camacho, R., …, Méndez, A. (2006). Increased cytogenetic damage in a zone in transition from agricultural to industrial use: comprehensive analysis of the micronucleus test in peripheral blood lymphocytes. *Mutagenesis.* 21 (5), 335–342.

Ocampo-Fletes, I., Parra-Inzunza, F., & Ruiz-Barbosa, Á. E. (2018). Derechos al uso de agua y estrategias de apropiación en la región semiárida de Puebla, México. *Agricultura, sociedad y desarrollo*, *15*(1), 63-83.

Pearce, D. W., Moran, D., & Krug, W. (1999). The global value of biological diversity: a report to UNEP. *Center for Social and Economic Research on the Global Environment-CSERGE, University College, Londres*.

Pérez Castresana, G., Tamariz Flores, V., López Reyes, L., Hernández Aldana, F., Castelán Vega, R., Morán Perales, J., ... & Handal Silva, A. (2018). Atoyac River pollution in the metropolitan area of Puebla, México. *Water*, *10*(3), 267.

Pérez Castresana, G., Castañeda Roldán, E., García Suastegui, W. A., Morán Perales, J. L., Cruz Montalvo, A., & Handal Silva, A. (2019). Evaluation of Health Risks Due to Heavy Metals in a Rural Population Exposed to Atoyac River Pollution in Puebla, Mexico. *Water*, *11*(2), 277.

Popa, F., & Guillermin, M. (2017). Reflexive methodological pluralism: the case of environmental valuation. *Journal of Mixed Methods Research*, *11*(1), 19-35.

Rodríguez, J. C. A., & Beristaín, B. T. (2014). Las Áreas Naturales Protegidas: entre la depredación y el uso sustentable. Área Natural Protegida Reserva Ecológica Natural de la Cuenca Alta del Río Atoyac. Mimeo.

Rodríguez, L.; Morales, J. & Zavala, P. (2012). Evaluación socioeconómica de daños ambientales por contaminación del Río Atoyac en México. *TyCA-RETAC*, 3, 143–151.

Rodríguez-Espinosa, P. F., Shruti, V. C., Jonathan, M. P., & Martinez-Tavera, E. (2018). Metal concentrations and their potential ecological risks in fluvial sediments of Atoyac River basin, Central Mexico: Volcanic and anthropogenic influences. *Ecotoxicology and Environmental Safety*, *148*, 1020-1033.

Rodríguez-Tapia, L., & Morales-Novelo, J. A. (2017). Bacterial pollution in river waters and gastrointestinal diseases. *International journal of environmental research and public health*, *14*(5), 479.

Rodríguez-Tapia, L., Morales-Novelo, J. A., & Zavala-Vargas, P. (2012). Evaluación socioeconómica de daños ambientales por contaminación del río Atoyac en México. *Tecnología y ciencias del agua*, *3*, 143-151.

Sandoval, A. M., Pulido, G., Monks, S., Gordillo, A. J. & Villegas, E. C. (2009). Evaluación fisicoquímica, microbiológica y toxicológica de la degradación ambiental del Río Atoyac, México. *Interciencia*, 34: 880-887.

Sierra, E. M., Pale, R. C. C. & Serrano, A. M. (2005). Análisis Legislativo y de Políticas Públicas en medio ambiente y salud en la Cuenca del Alto Río Atoyac. Centro Fray Julián Garcés de Derechos Humanos y Desarrollo Local A.C.

Sistema Estatal de Información (2015). Información estadística general de Municipios del estado de Puebla. Available online: http://datos.puebla.gob.mx/dataset/informacion-estadistica-general-municipios-estado-puebla (Accessed on 4 Dec 2020)

Stiglitz, J. E., & Rosengard, J. K. (2015). *Economics of the public sector: Fourth international student edition*. WW Norton & Company.

UN Water (2013). Water security and the global water agenda: a UN-water analytical brief. *Hamilton, ON: UN University*.

Urrútia, G., & Bonfill, X. (2010). Declaración PRISMA una propuesta para mejorar la publicación de revisiones sistemáticas y metaanálisis. *Medicina Clinica*, *135*(11), 507–511. http://dx.doi.org/10.1016/j.medcli.2010.01.015